\documentclass[aps,prl,twocolumn,showpacs,,superscriptaddress,groupedaddress]{revtex4}

\usepackage{times,mathptmx,amssymb} \usepackage{graphicx}
\usepackage{amsmath} 
\usepackage{subfigure}
\usepackage{epsfig}
\usepackage{graphicx}

\usepackage{color}
\usepackage{soul}
\usepackage{epstopdf}

\begin{document}
\newcommand{\Ant}[1]{{#1}}
\newcommand{\Delete}[1]{}
\newcommand{\kcomment}[1]{#1}

\title{Shaping Segregation: Convexity vs. concavity}

\author{S. Gonz\'alez}
\affiliation{Multi-Scale Mechanics, Dept. of Mechanical Engineering, MESA+, University of Twente, P.O. Box 217, 7500 AE Enschede,
  The Netherlands}
\author{C. R. K Windows-Yule}
\affiliation{School of Physics and Astronomy, University of Birmingham, UK, B15 2TT}
\author{ S. Luding}
\affiliation{Multi-Scale Mechanics, Dept. of Mechanical Engineering, MESA+, University of Twente, P.O. Box 217, 7500 AE Enschede,
  The Netherlands}
\author{D. J.Parker}
\affiliation{School of Physics and Astronomy, University of Birmingham, UK, B15 2TT}
\author{A.R. Thornton$^{1,}$}
\thanks{\mbox{a.r.thornton@utwente.nl}}

\affiliation{Mathematics of Computational Science, Dept. of
  Appl. Mathematics, University of Twente, P.O. Box 217, 7500 AE Enschede,
  The Netherlands}

\begin{abstract}
Controlling segregation is both a practical and a theoretical challenge. In this Letter we demonstrate a manner in which rotation-induced segregation may be controlled by altering the geometry of the rotating containers in which granular systems are housed. Using a novel drum design comprising concave and convex geometry, we explore a means by which radial size-segregation may be used to drive axial segregation, resulting in an order of magnitude increase in the axial segregation rate. This finding, and the explanations provided of its underlying mechanisms, could lead to radical new designs for a broad range of particle processing applications.
\end{abstract}

\maketitle

Granular flows in rotating drums are widely used to study mixing, segregation, and pattern formation \cite{seiden11}.  While the most studies focus on a circular cylinder geometry, several recent works have explored different drum geometries \cite{Hill2001,Meier2006,MeierLueptowOttino2007,Naji2009,christov2010,PrasadKhakhar2010,Pohlman2012}. {Non-circular drums are important from both an application perspective, as they are used in various industries, and from a theoretical perspective, to further validate theoretical approaches developed primarily from simpler cylindrical configurations.} Segregation is known to occur in rotated granular materials. Radial segregation occurs after a few rotations \cite{Hill2001} and, if rotation continues for an adequately long duration, axial segregation may also appear \cite{zik1994}. The question arises: is it possible to control this axial segregation? 

The role of geometry in segregation is well known \cite{Metcalfe1995}, with experiments looking at a variety of convex \cite{KawaguchiTsutsumiTsuji2006,Naji2009} and concave \cite{Cleary2003,Morton2004,GuptaKatterfeldSoetemanLuding2010,PrasadKhakhar2010} drums. However, to the best of our knowledge this is the first study utilising mixed convex-concave systems. In geometry concave polygons are defined by two numbers $\{n/m\}$; $n$ is the number of sides (points) and the polygon is formed by connecting every $m^{{th}}$ point with straight lines. We chose the simplest regular concave polygon the $\{5/2\}$-star polygon, or pentagram. In this Letter, we first focus on the dynamics of mono-sized particles in simple concave drums and later explain how the novel combination of concave and convex shapes can be used to control segregation.

{\em Experimental set-up.} Drums of length $a = 119mm$ and width $\Delta z \in (10,24)$ mm are partially filled with glass beads of diameter $d = 3.5 \pm 0.3$ and rotated at a constant rate $\Omega = \pi/2$ rad/s. \kcomment{Data is acquired from the experimental system using both optical techniques and positron emission particle tracking (PEPT), enabling the bed's exterior and interior to be explored. PEPT is a non-intrusive technique which records the motion of a single `tracer' particle in order to extrapolate of a variety of time-averaged quantities pertaining to the system as a whole. Although not necessary to the understanding of this Letter, for the interested reader, a comprehensive overview of the PEPT technique may be found in our references \cite{parker2002positron,wildmanSingle}.}

In simulations, experimental system dimensions and particle properties are used, with particles' contact forces represented using a standard spring dash-pot model \cite{cundall79,Luding2008a,Luding2008b} as implemented in \cite{thornton2012}. The drum rotation is achieved by changing the direction of gravity at fixed $\Omega = \pi/2$ rad/s in order to consistently provide a continuous free-surface avalanche \cite{Mellmann2001}. \kcomment{Simulations are conducted with both periodic boundary conditions in the axial direction, as well as with solid \lq side-walls\rq\, allowing both direct comparison with experimental results in the case of the convex-concave drum, and investigation of the effect of drum geometry on flow for purely convex or concave systems using periodic walls.
}

\begin{figure}[!ht]
  \centering

  \epsfig{file=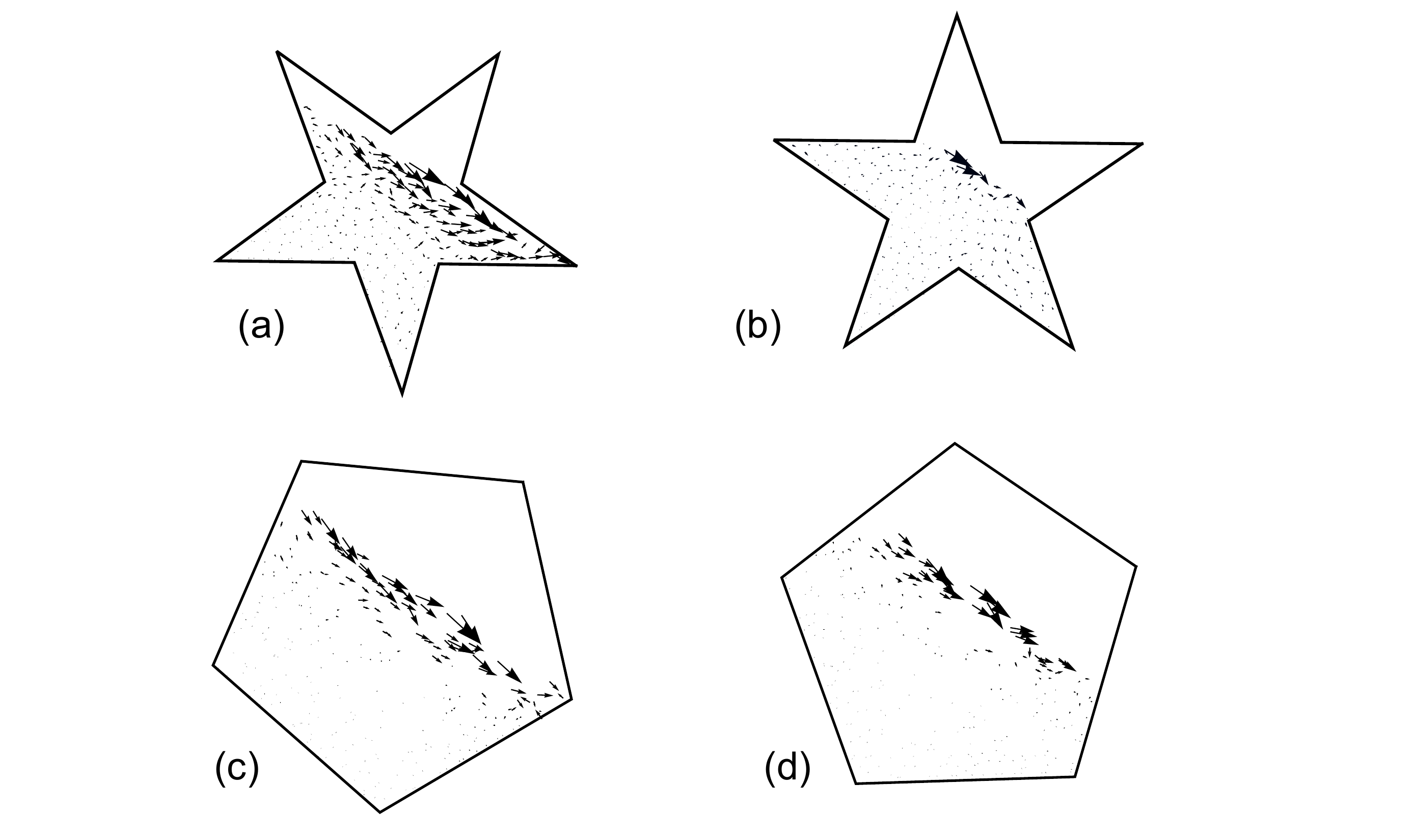,width=.9\columnwidth}

  \caption{Instantaneous velocity fields for $F = 60\%$, $\Delta z = 10$ mm. Arrows represent velocity, $v$, projected on the $x$-$y$ plane for all particles. 
  }
  \label{fig2}
\end{figure}

\begin{figure}[ht!]
  \centering
  \epsfig{file=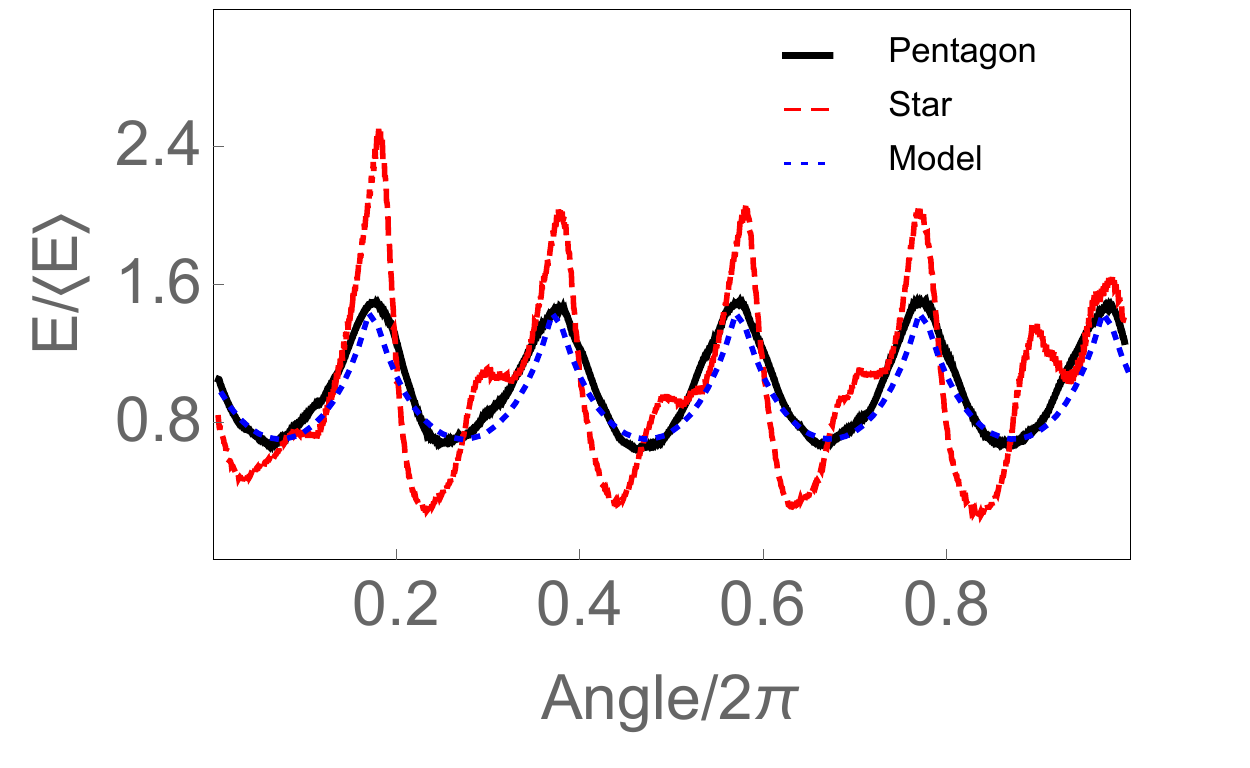,width=.7\columnwidth}  

\caption{(Colour online) Kinetic energy vs.\ time during   one cycle for a simulated pentagonal drum (solid, black), a pentagram (red, dashed), and the model for the pentagon (blue,
  dotted). 
  }  \label{fig3}
\end{figure} 

{\em The effect of the filling fraction.} We observe four differing flow regimes depending on $F$. If grains occupy a volume smaller than one leg of the pentagram,
they flow intermittently from leg to leg. If the $F$ increases such that there are always grains in at least two legs, flow is constant but its angle changes continuously. When two to four legs are filled, flow is continuous but with two qualitatively different flow profiles depending on the drum's angle. Once grains occupy more than four legs, flow becomes intermittent and grain displacement is strongly limited, decreasing transport in the bulk, with dynamics mostly due to geometrical rearrangements. We focus on the regime $40\% \leq F \leq 60\%$, where unsteady flow is produced by a concave drum rotating at a constant rate. The geometric shape naturally causes periodic changes in the flowing layer as a function of the instantaneous orientation of the pentagram, as recently reported in other geometries \cite{Pohlman2012}.

{\em Comparison of Pentagram and Pentagon.}  In cylindrical and general convex drums, steady flow has a roughly constant kinetic energy, $E$, independent of the angle of rotation. For the pentagram, however, this oscillates strongly: when a pentagram points up, flow is slow while when pointing down, flow is much faster. Fig.\ \ref{fig2} shows velocity fields for both pentagram (a, b) and pentagon (c, d). The pentagram shows great variation in the magnitude of $v$ between the up (b) and down configuration (a): when the pentagram points down, the avalanche occurs in a thick layer. As the drum rotates, more space becomes available, producing a saltating flow. This creates a fast avalanche in the down part of the flow, and the consequent movement of all the flowing layer. Thus, $E$ shows five maxima during a cycle (Fig.\ \ref{fig3}). By allowing particles more space to flow, a large, fast, avalanche is produced. This avalanche is not symmetric along the free-surface. Most of the kinetic energy is on the downside, where the free volume makes it easier to flow. Eventually, the leg is filled with particles and the avalanche recovers its slow flow, before the process repeats. We now focus on how this feature can be used to control segregation. To do this, one must introduce the pentagram's convex counterpart, the pentagon. As the pentagon rotates, the total length of the flowing layer changes, creating an oscillation in $E$ with the same period as for the pentagram (see Fig.\ \ref{fig3}). However, this flow, and its velocity, are much more consistent in the pentagon, with a smaller variation between minimum and maximum. 

The periodic structure of the $E$ can be understood by simple arguments. If one considers the speed of the flowing layer and its depth constant as much smaller than the filling height, $H$, then $E$ is proportional to the length, $L$, of the flowing layer \ \cite{Pohlman2012}. Disregarding the angle of the walls, and assuming a straight free-surface, $L$ scales approximately as $L \propto 1/\cos(\theta)$, with $\theta \in [0,2\pi/5]$ the angle of rotation modulo the shape's symmetry, in this case $2\pi/5$. Hence, $E\propto 1/\cos(\theta)^2$. The agreement of simulations with this simple model is remarkable (see Fig.\ \ref{fig3}), \kcomment{although deviations from this simple sinusoidal form arise for the concave drum. These deviations are not surprising, as while in a pentagon the flowing layer is at the edge of the geometric region of constant volume and this region is always connected, giving a relatively consistent filling fraction, $F$, the same does not hold for pentagrams and other concave shapes. Consequently, both of the above assumptions are likely to be broken for such geometries. Specifically, one observes the presence of local maxima preceding each of the main peaks in kinetic energy. It is also notable that the initial maximum in $E$ is markedly higher than the following peaks. The former of these deviations can be explained by the fact that particles in the lower region of the surface flow avalanche first over the lowermost leg, before being followed by grains in the middle and upper regions, thus leading to the observed `two-part' increase in $E$ \footnote{See the supplementary video available at \hl{XXX}}. The latter, meanwhile, can be explained by the initial presence of localised jamming within the system, whereby a collection of particles in a jammed state are able to reach a higher point in the system before avalanching, naturally resulting in a higher-than-average kinetic energy. It is finally worth noting that the existence of side-walls acts to frustrate the observed local maxima, while the general sinusoidal evolution of $E$ is found to persist - i.e. the concave system approaches more closely the theoretical form.}

\begin{figure}[ht!]
  \centering
\epsfig{file=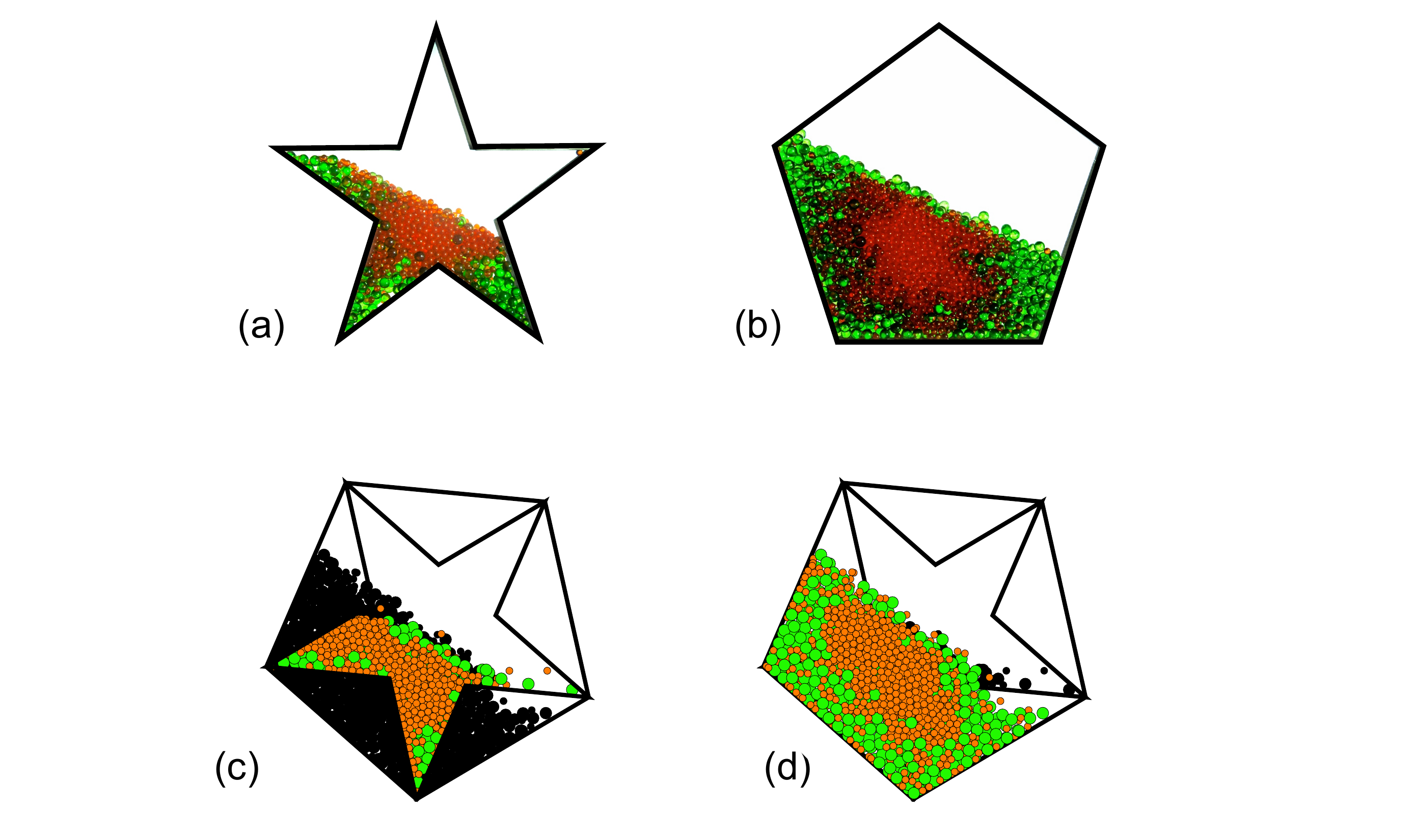,width=.9\columnwidth} 
\caption{ (Colour online) Photographs of the experiment after four revolutions in the
  axially homogeneous drum for the pentagram (a) and the pentagon
  (b). Simulations for the axially inhomogeneous, layered drum
  after four revolutions, from the pentagram-shaped
  side (c), and from the  pentagonal side (d). Particles are
  coloured by size with orange small and green big; black particles
  correspond to those particles that belong to the opposite side of
  the drum.
   } \label{fig4}
\end{figure}

\begin{figure}[ht!]
  \centering
\epsfig{file=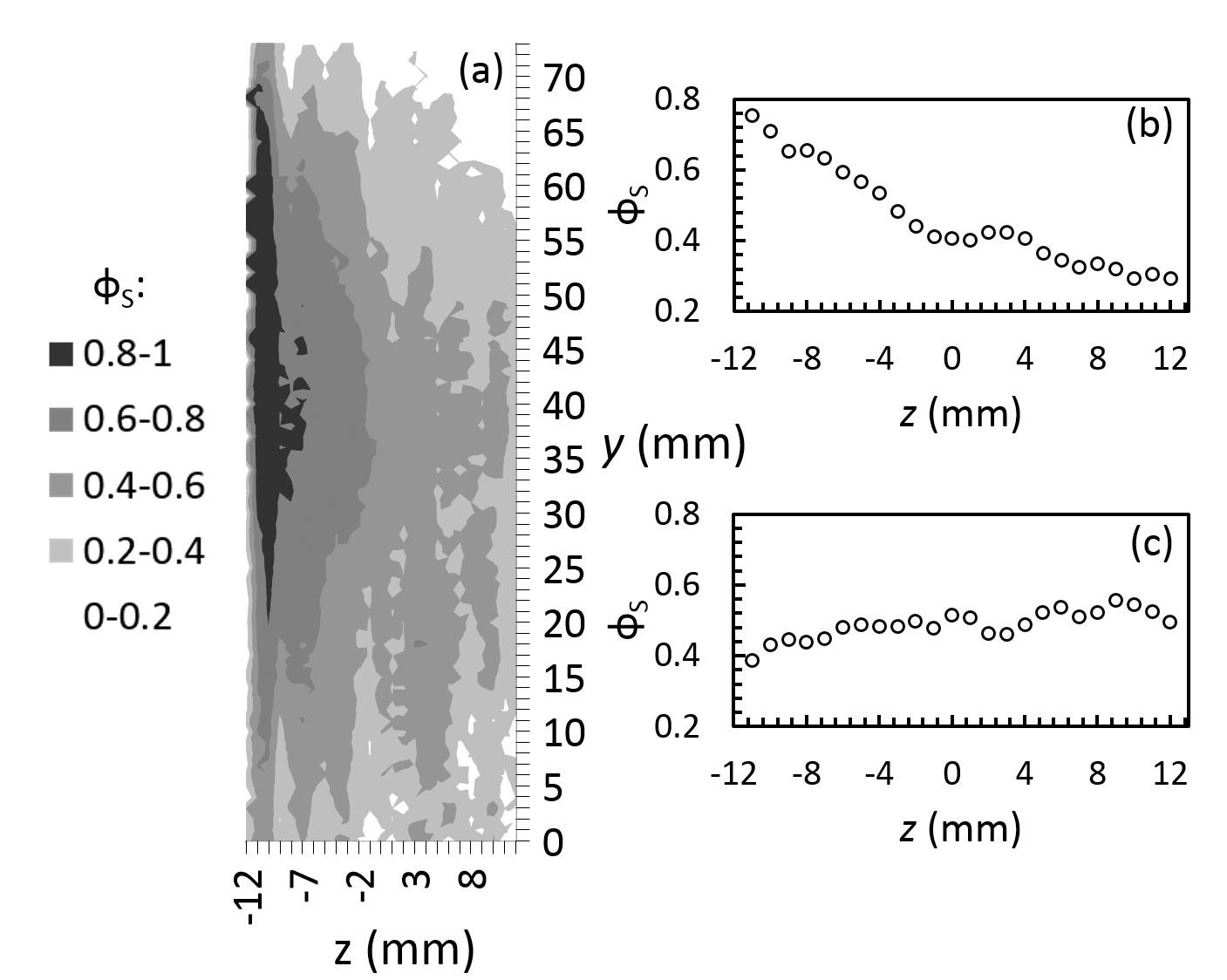,width=.75\columnwidth} 
\caption{\kcomment{Experimental data acquired using PEPT showing the time-averaged spatial variation of the fractional concentration of small particles, $\phi_S$, for a two-dimensional (2D) projection through the $x$-axis, and as a 1D profile along the axial ($z$) direction. In panels (a) and (b), negative $z$ represents the concave, and positive $z$ the convex sides of a pentagram-pentagon drum. Shown also is the 1D profile for an equivalent, purely convex system (c), illustrating the enhanced segregation produced by the dual-geometry system. In all cases, $F = 60 \%$, $\Delta z = 24$ mm.}
   } \label{peptSeg}
\end{figure}

{\em Shape-induced axial segregation.} Although the influence of container shape on segregation has already been reported \cite{Mao-Bin2005,hu2005} this is the first time that it is
used in a rotating drum. It is also known that modifying the geometry (e.g.\ adding obstacles or mixing blades) can \emph{reduce} segregation \cite{Shi2007} but it has not been shown how to \emph{augment} and \emph{control} it -- a matter of obvious practical importance. For bi-disperse granulates in any rotating container, small particles will migrate towards the drum's centre \cite{Mounty2007,seiden11} (see Fig.\ \ref{fig4} (a) and (b)). For adequately wide systems, upon continuous rotation the system will segregate axially \cite{zik1994}, a process orders of magnitude slower than radial segregation. However, if two different geometries are used along the axial direction, e.g.\ a half-pentagram, half-pentagon drum, the usually slow segregation along this axis can be enhanced and its direction controlled (see Fig.\ \ref{fig4} (c) and (d)). In both experiment and simulation, two sections of equal width are combined. We use particles of $d = 4.0, 2.5$ mm, in an equal volume distribution. The rapid axial segregation happens only with a convex-concave combination, as for convex shapes there exists little difference in the level of the flow, just the length of avalanching layer \cite{ArntzOttherBeeftinkBoomBriels2013}. We performed several experiments, putting together circular and square sections, pentagonal and square, and differently oriented square sections. None of these configurations presented axial segregation on the time scale of observation ($\sim 20$ revolutions). \kcomment{A clearer representation of the segregation -- both axial and radial -- achieved in the convex-concave system described above may be seen in Fig. \ref{peptSeg}. It should be noted that, due to the somewhat constrained nature of the system under investigation, the degree of axial segregation observed is likely to be \emph{lower} than it would be in comparatively longer drums, due to the restricted motion of particles in the axial direction. Thus, the significant segregation observed even in these unfavourable conditions is a highly pleasing result, as one may well expect larger systems to provide still greater separation}.

Grains tend to minimise their potential energy, i.e.\ move towards the concave side, which also possesses more free-volume. Note that the few large grains in the run-out leg of Fig.\ \ref{fig4} (c) will eventually fall to the pentagonal side. Radial segregation occurs in each side, so large particles go to the surface and small to the centre. Since for this packing fraction the avalanche in the pentagonal side of the drum ($z<0$) is slower than in the pentagram-shaped section ($z>0$), large particles can move to the empty side since they are faster and there is space available for them. Once the two avalanches reach the same angle there is no more flux of particles. This process is repeated five times per revolution (see Fig.\ \ref{fig:CMz} (a)). When the big particles drop from the run-out leg of the pentagram to the pentagon, the centre of mass of the large particles shifts towards the pentagon. However, the process is not completely irreversible; some, but fewer, large particles go again to the pentagram side as the drum rotates. In this way, an oscillating movement of the centre of mass of each species is observed: big particles fall to the pentagonal side when the run-out leg is empty; once the flow covers the run-out leg some large particles return to the pentagram side. By this mechanism, there is a net transport of large particles to the convex side of the drum, while the concave side becomes dominated by small particles. Since this mechanism relies on the
fast radial segregation, it is orders of magnitude faster than the axial segregation previously reported for axially homogeneous drums \cite{zik1994}.

\begin{figure}[ht!]
  \centering
  \scalebox{1}{\includegraphics[width=.75\columnwidth]{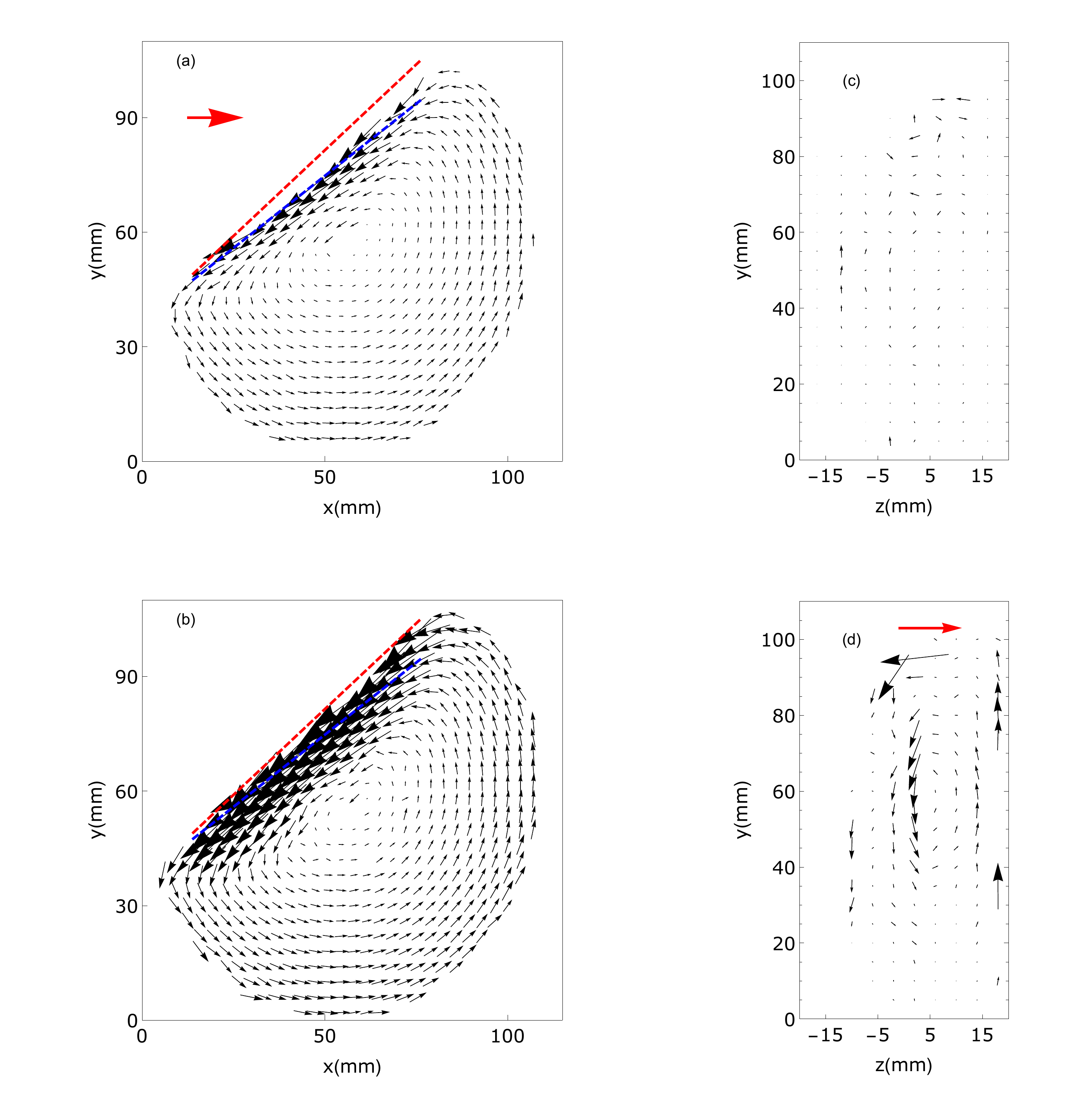}}

\caption{\kcomment{Time-averaged velocity fields for a pentagram-pentagon system of width $\Delta z = 24$ mm. Panels (a) and (b) show, for the convex and concave regions respectively, the radial velocities of particles in the cartesian $x$-$y$ plane, where the $x$ and $y$ directions lie perpendicular to the axial $z$ direction in the vertical ($y$) and horizontal ($x$) planes. 
Differences in angle between the free surfaces of the bed for the two regions are emphasised through the inclusion of a red dashed line corresponding to the concave region, and a blue dotted line representing the convex region. 
Panels (c) and (d) show depth-averaged particle flow in the $z$-$y$ plane for small (panel (c)) and large (panel (d)) particles belonging to the same system.  In all images presented, the direction and magnitude of the average particle velocity in a given region of the experimental volume are represented by the orientation and length of the arrows displayed.
   } } \label{peptV}
\end{figure}

\begin{figure}[ht]
  \centering

 \epsfig{file=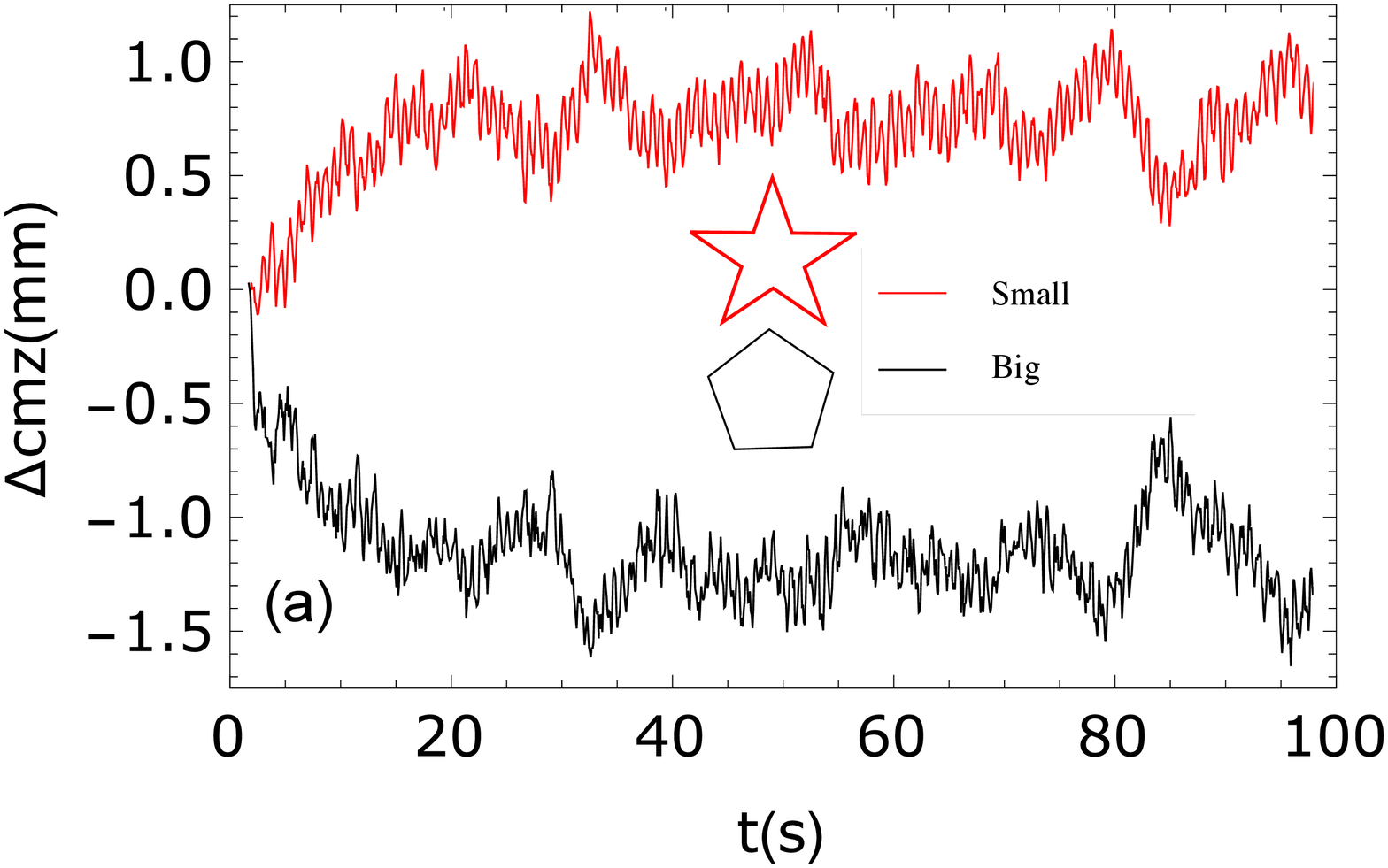,width=0.45\columnwidth}
\hspace{.5cm}
\raisebox{.285cm}{\epsfig{file=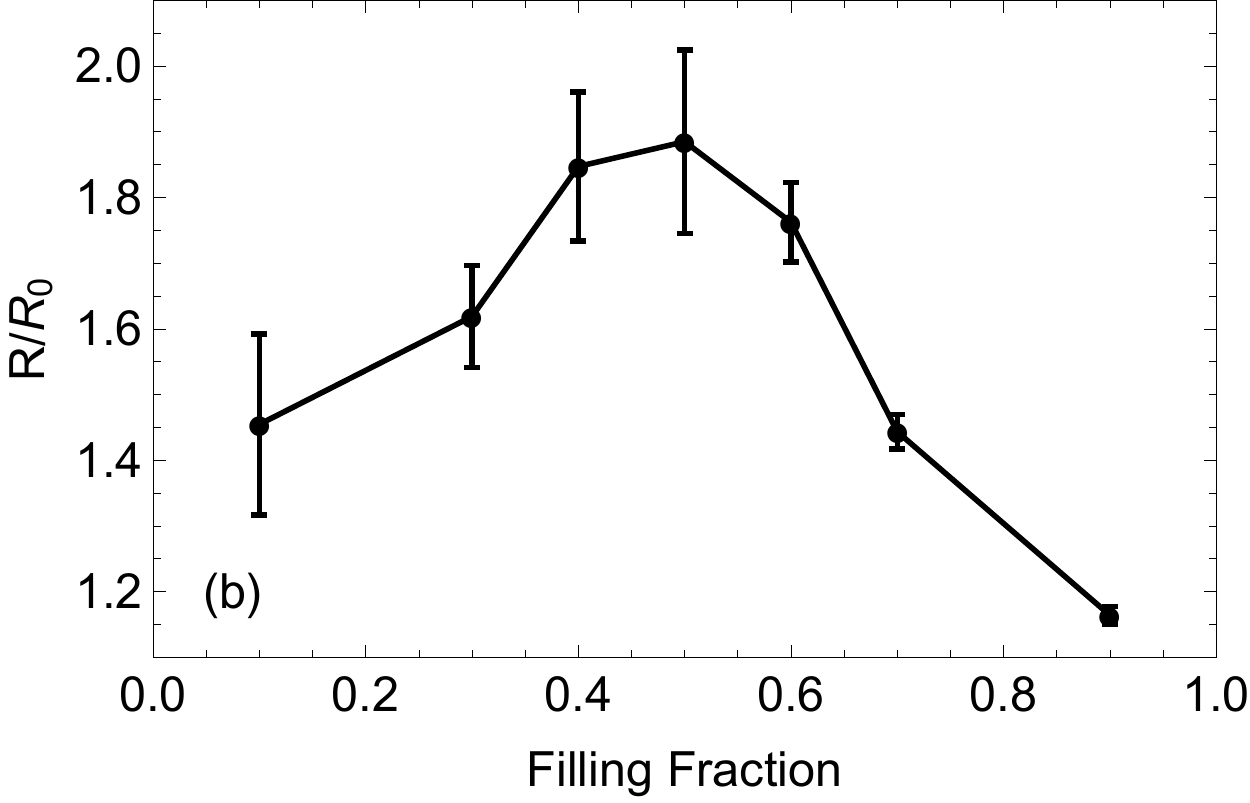,width=0.4\columnwidth}}
\caption{(Colour online) Left: evolution of the displacement of the centre of mass for simulations at $F = 50\%$ in a pentagram-pentagon geometry of width $\Delta z = 22$ mm. Right: Number ratio ($R = N_{\rm big}/N_{\rm small}$) in the pentagonal side of the drum ($z<0$) versus the filling fraction normalised by the initial conditions. Data is averaged in ten snapshots during two turns of the drum. Error bars are the standard deviation of these measurements.}
\label{fig:CMz}
\end{figure}

\kcomment{Experimental evidence of the mechanism proposed above may be seen in Fig.\ \ref{peptV}. From these images it is clear that, as expected, there exists a difference in the level and angle of inclination of the bed's surface between the convex and concave sides of the drum. Moreover, the velocity fields in panels (a) and (b) demonstrate the avalanching region on the bed's concave side to be both faster and deeper than for the convex half, again in agreement with our hypothesis. Panels (c) and (d) show two-dimensional, depth averaged velocity fields for the $z$-$y$ plane, where $y$ denotes a vertical axis perpendicular to the axial ($z$) axis. In image (d), which corresponds to the time-averaged motion of the large particles in the system, we see evidence of the recirculatory transport discussed above, and whose effect on the system's mass centre is shown in  Fig.\ \ref{fig:CMz} (a). It is interesting to note that such motion is seemingly absent for the small particles in the same system (panel (c)).}

Finally, it must be noted that the final degree of segregation is not the equal every $F$. Fig.\ \ref{fig:CMz} (b) shows the change in the number of large particles in the pentagonal side of the drum for different filling fractions. If $F$ too low, the avalanche on the pentagram side of the drum arrives concurrently with the one in the pentagon and axial segregation is slower. One could argue, that excluded volume effects make the small particles go preferably to the pentagram side since the big particles do not fit into the legs so easily, as reported in \cite{hu2005}. However, this mechanism alone does not explain the maximum in segregation at $\sim 50\%$ filling fraction. This can only be due to the differential flows along the axial direction and the consequent conversion of radial to axial segregation previously discussed.

{\em Conclusion.} In this letter we have studied granular flows inside the simplest possible regular concave drum, that is, the pentagram-drum. Different regimes are found for a fixed angular velocity depending on the filling fraction. From intermittent avalanching (low filling fraction) to geometrical rearrangements (high filling) passing by continuous flow (intermediate filling). These flow patterns differ qualitatively from those observed in convex drums. We have used this insight to control the segregation of a binary granular system, achieving fast geometrically-induced size separation along the axial direction, by re-directing radial segregation into axial segregation. The possible applications of this mechanism are potentially revolutionary. Firstly, one may produce axial segregation orders of magnitude faster than previously reported. Secondly, the direction and rate of  segregation can be controlled. Finally, it makes us reconsider the role of boundary conditions when dealing with granular materials; this is the first step in shaping segregation at our will. 

The practical importance of this discovery can be far-reaching in industries ranging from pharmaceuticals to mining. We foresee several applications for our discovery including, but by no means limited to, rotating kilns -- allowing differential residence times depending on the size of the particles -- and milling devices -- whereby creating a sandwich of concave sections with a convex shape in the middle, large particles can be conducted into the middle of the mill, thus increasing efficiency by keeping the grinders and larger particles in the mill while moving the fines to the ends, where they could be removed. \kcomment{ This study also provides great scope for future work in the extension, refinement and practical application of the findings presented here.}

We thank W.\ Zweers for the construction of the experimental set-up;
V.\ Ogarko for generating the bi-disperse packings; W. den Otter and
N. Rivas for the critical reading of the manuscript. The simulations
performed for this paper are undertaken with \url{MercuryDPM.org}; 
primarily developed by T. Weinhart, A. R. Thornton and D. Krijgsman at
the University of Twente. This study was supported by the Stichting
voor Fundamenteel Onderzoek der Materie (FOM), financially supported
by the Nederlandse Organisatie voor Wetenschappelijk Onderzoek (NWO),
through the FOM project 07PGM27.

\bibliographystyle{unsrt}

\bibliography{new200409,granulates,biblio,paper}

\begin{thebibliography}{10}

\bibitem{seiden11}
G.~Seiden and P.~J. Thomas.
\newblock Complexity, segregation, and pattern formation in rotating-drum
  flows.
\newblock {\em Rev. Mod. Phys.}, 83:1323--1365, Nov 2011.

\bibitem{Hill2001}
K.~M. Hill, N.~Jain, and J.~M. Ottino.
\newblock Modes of granular segregation in a noncircular rotating cylinder.
\newblock {\em Phys. Rev. E}, 64:011302, Jun 2001.

\bibitem{Meier2006}
S.~W. Meier, S.~E. Cisar, R.~M. Lueptow, and J.~M. Ottino.
\newblock Capturing patterns and symmetries in chaotic granular flow.
\newblock {\em Phys. Rev. E}, 74:031310, Sep 2006.

\bibitem{MeierLueptowOttino2007}
S.~W. Meier, R.~M. Lueptow, and J.~M. Ottino.
\newblock A dynamical systems approach to mixing and segregation of granular
  materials in tumblers.
\newblock {\em Adv. In. Phys.}, 56(5):757--827, 2007.

\bibitem{Naji2009}
L.~Naji and R.~Stannarius.
\newblock Axial and radial segregation of granular mixtures in a rotating
  spherical container.
\newblock {\em Phys. Rev. E}, 79:031307, Mar 2009.

\bibitem{christov2010}
I.~C. Christov, J.~M. Ottino, and R.~M. Lueptow.
\newblock Chaotic mixing via streamline jumping in quasi-two-dimensional
  tumbled granular flows.
\newblock {\em Chaos: An Interdisciplinary Journal of Nonlinear Science},
  20(2):023102, 2010.

\bibitem{PrasadKhakhar2010}
D.~V.~N. Prasad and D.~V. Khakhar.
\newblock Mixing of granular material in rotating cylinders with noncircular
  cross-sections.
\newblock {\em Physics of Fluids}, 22(10):103302, 2010.

\bibitem{Pohlman2012}
N.~A. Pohlman and D.~F. Paprocki~Jr.
\newblock Transient behavior of granular materials as result of tumbler shape
  and orientation effects.
\newblock {\em Granular Matter}, pages 1--9, 2012.

\bibitem{zik1994}
O.~Zik, D.~Levine, S.~G. Lipson, S.~Shtrikman, and J.~Stavans.
\newblock Rotationally induced segregation of granular materials.
\newblock {\em Phys. Rev. Lett.}, 73:644--647, Aug 1994.

\bibitem{Metcalfe1995}
G.~Metcalfe, T.~Shinbrot, J.~J. McCarthy, and J.~M. Ottino.
\newblock Avalanche mixing of granular solids.
\newblock {\em Nature}, 374(6517):39--41, Mar 1995.

\bibitem{KawaguchiTsutsumiTsuji2006}
Toshihiro Kawaguchi, Kenji Tsutsumi, and Yutaka Tsuji.
\newblock Mri measurement of granular motion in a rotating drum.
\newblock {\em Particle \& Particle Systems Characterization},
  23(3-4):266--271, 2006.

\bibitem{Cleary2003}
P.~W. Cleary, R.~Morrisson, and S.~Morrell.
\newblock Comparison of {DEM} and experiment for a scale model {SAG} mill.
\newblock {\em International Journal of Mineral Processing}, 68(1–4):129 --
  165, 2003.

\bibitem{Morton2004}
D.~Morton and S.~Dunstall.
\newblock Using the {W}eb to increase the availability of {DEM}-based mill
  modelling.
\newblock {\em Minerals Engineering}, 17(11–12):1199 -- 1207, 2004.

\bibitem{GuptaKatterfeldSoetemanLuding2010}
Akash Gupta, Andr{\'e} Katterfeld, Bastiaan Soeteman, and Stefan Luding.
\newblock Discrete element study mixing in an industrial sized mixer.
\newblock {\em World Congress Particle Technology 6, Nuremberg,
  CD-Proceedings}, 2010.

\bibitem{parker2002positron}
DJ~Parker, RN~Forster, P~Fowles, and PS~Takhar.
\newblock Positron emission particle tracking using the new birmingham positron
  camera.
\newblock {\em Nuclear Instruments and Methods in Physics Research Section A:
  Accelerators, Spectrometers, Detectors and Associated Equipment},
  477(1):540--545, 2002.

\bibitem{wildmanSingle}
RD~Wildman, JM~Huntley, J-P Hansen, DJ~Parker, and DA~Allen.
\newblock Single-particle motion in three-dimensional vibrofluidized granular
  beds.
\newblock {\em Physical Review E}, 62(3):3826, 2000.

\bibitem{cundall79}
P.~A. Cundall and O.~D.~L. Strack.
\newblock A discrete numerical model for granular assemblies.
\newblock {\em G\'eotechnique}, 29(1):47--65, 1979.

\bibitem{Luding2008a}
S.~Luding.
\newblock Introduction to discrete element methods {DEM}: Basics of contact
  force models and how to perform the micro-macro transition to continuum
  theory.
\newblock {\em Euro. J. of Enviro. Civ. Eng.}, 12(7-8):785--826, 2008.

\bibitem{Luding2008b}
S.~Luding.
\newblock The effect of friction on wide shear bands.
\newblock {\em Particulate Science Technology}, 26:33--42, 2008.

\bibitem{thornton2012}
A.~R. Thornton, T.~Weinhart, S.~Luding, and O.~Bokhove.
\newblock Modeling of particle size segregation: Calibration using the discrete
  particle method.
\newblock {\em International Journal of Modern Physics C}, 23(08):1240014,
  2012.

\bibitem{Mellmann2001}
J.~Mellmann.
\newblock The transverse motion of solids in rotating cylinders‚Äîforms of
  motion and transition behavior.
\newblock {\em Powder Technology}, 118(3):251 -- 270, 2001.

\bibitem{Mao-Bin2005}
M.~B. Hu, X.~Z. Kong, Q.~S. Wu, and Y.~H. Wu.
\newblock Effects of container geometry on granular segregation pattern.
\newblock {\em Chinese Physics B}, 14(9):1844, 2005.

\bibitem{hu2005}
M.~B. Hu, X.~Z. Kong, Q.~S. Wu, and Y.~H. Wu.
\newblock Granular segregation in a multi-botleneck container: Mobility effect.
\newblock {\em International Journal of Modern Physics B}, 19(10):1793--1800,
  2005.

\bibitem{Shi2007}
Deliang Shi, Adetola~A. Abatan, Watson~L. Vargas, and J.~J. McCarthy.
\newblock Eliminating segregation in free-surface flows of particles.
\newblock {\em Phys. Rev. Lett.}, 99:148001, Oct 2007.

\bibitem{Mounty2007}
D.~Mounty.
\newblock {\em Particle Size Segregation in Convex Rotating Drums}.
\newblock PhD thesis, University of Manchester, 2007.

\bibitem{ArntzOttherBeeftinkBoomBriels2013}
M.M.H.D. Arntz, W.K. den Otter, H.H. Beeftink, R.M. Boom, and W.J. Briels.
\newblock The influence of end walls on the segregation pattern in a horizontal
  rotating drum.
\newblock {\em Granular Matter}, 15(1):25--38, 2013.

\end{thebibliography}

\end{document}